\documentclass[useAMS,usenatbib]{mn2e}
\usepackage{psfig}

\usepackage{times}
\def\Liso{\mbox{$L_{\rm iso}$}}

\def\Ep{\mbox{$E_{\rm pk}$}}

\def\tdur{\mbox{$T_{0.45}$}}
\def\tdurob{\mbox{$T_{0.45}^{\rm obs}$}}

\def\relL{\mbox{\Liso--\Ep--\tdur}}
\def\chs{\mbox{$\chi^2$}}

\def\chsr{\mbox{$\chi_r^2$}}

\def\Om{\mbox{$\Omega_{\rm m}$}}
\def\OL{\mbox{$\Omega_{\Lambda}$}}

\def\Ot{\mbox{$\Omega_{\rm tot}$}}
\def\rde{\mbox{$\rho_{\rm DE}$}}
\def\wz{\mbox{$w(z)$}}
\def\wo{\mbox{$w_0$}}
\def\wu{\mbox{$w_1$}}

\def\zt{\mbox{$z_t$}}

\def\spose#1{\hbox to 0pt{#1\hss}}
\newcommand\lsim{\mathrel{\spose{\lower 3pt\hbox{$\mathchar"218$}}
     \raise 2.0pt\hbox{$\mathchar"13C$}}}
\newcommand\gsim{\mathrel{\spose{\lower 3pt\hbox{$\mathchar"218$}}
     \raise 2.0pt\hbox{$\mathchar"13E$}}}

\title[The Hubble diagram at $z>>1$]
{The Hubble diagram extended to $z>>1$: the $\gamma-$ray properties of GRBs 
confirm the $\Lambda$CDM model}

\author[Firmani et al.]
{C. Firmani$^{1,2}$\thanks{E--mail: firmani@merate.mi.astro.it},
V. Avila--Reese$^{2}$, G. Ghisellini$^{1}$ and G. Ghirlanda$^{1}$\\
$^{1}$Osservatorio Astronomico di Brera, via E.Bianchi 46, I-23807
Merate, Italy\\
$^{2}$Instituto de Astronom\'{\i}a, Universidad Nacional Aut\'onoma de M\'exico,
A.P. 70-264, 04510, M\'exico, D.F.}

\begin{document}

% \date{Accepted 1988 December 15. Received 1988 December 14; 
% in original form 1988 October 11}

\pagerange{\pageref{firstpage}--\pageref{lastpage}} \pubyear{2002}

\maketitle

\label{firstpage}

\begin{abstract}
Tight constraints on cosmological parameters can be obtained with
standard candles spanning a range of redshifts as large as possible.
We propose to treat SNIa and long Gamma--Ray Bursts (GRBs) as a
single class of candles.  Taking advantage of the recent release of
the Supernova Legacy Survey and {\it the recent finding of a tight
correlation among the energetics and other prompt $\gamma-$ray
emission properties of GRBs}, we are able to standardize the
luminosities/energetics of both classes of objects. In this way
we can jointly use GRB and SNIa as cosmological probes 
to constrain \Om\ and \OL\ and the parameters of the Dark Energy 
equation of state through the same Bayesian method that we have, 
so far, applied to GRBs alone.
Despite the large disparity in number (115 SN
Ia versus 19 GRBs) we show that the constraints on \Om\ and \OL\ are
greatly improved by the inclusion of GRBs.  More
importantly, the result of the combined sample is in excellent
agreement with the $\Lambda$CDM concordance cosmological model and does 
not require an evolving equation of state for the Dark Energy.
\end{abstract}

%long and gamma-ray


%======================================================
\section{Introduction}
%======================================================

Cosmology passed from being mostly a theoretical science 
to be one of the most accurate physical sciences in the phenomenological sense. 
The recent high--precision measurements of cosmological parameters together 
with the spectacular advances in the understanding of 
cosmic structure formation, produced a coherent picture of the evolution of 
the Universe but, on the other hand, prompted new fundamental questions.
One of them is related to the expansion history of the Universe and the 
possibility that a repulsive medium with negative pressure (Dark Energy, 
hereafter DE) dominates its content.  The Hubble diagram 
(distance--redshift relation, HD hereafter)
provides a key probe of the Universe expansion history. Because of their
almost homogeneous intrinsic energetics, Type Ia Supernovae 
(SNIa) have been used as the main distance indicators for constructing the 
HD. However, their current observability is limited to redshifts 
$z\la 1.7$, and the high--$z$ SNIa could suffer intergalactic dust extinction 
and evolutive effects \citep[e.g.,][and the references therein]{ostman05}. 
Thus, the finding of alternative
and complementary cosmological distance indicators is highly
desirable. Gamma--ray Bursts (GRBs), after ``standardizing'' their
energetics through adequate relations, probed to be reliable distance
indicators detectable up to very high $z'$s and free from dust 
extinction. The recent discovery of a
tight correlation among prompt $\gamma-$ray emission observables 
{\it alone} \citep{paper1} opens new perspectives for the HD
method, where GRBs can be combined with SNIa to extend the HD 
to $z\sim 5-10$ \citep[][]{schaefer}.

The observations of SNIa demonstrated that the expansion of the Universe is 
accelerating \citep{riess98,perl99}. The main explanation to this
acceleration is the dominion of DE in the current cosmological dynamics, 
though departures from conventional physics, like modified 
gravity theories, are also considered. 
DE is characterized mainly by its equation-of-state
parameter, $w = p_{\rm DE}/\rde$, where $p_{\rm DE}$ and \rde\ are
the pressure and energy density of DE. For $w<-1/3$ the universe
undergoes accelerated expansion.  The simplest interpretation of DE is
the cosmological constant $\Lambda$, in which case $w=-1$ and
$\rde=\rho_{\Lambda}=$const. However, more exotic models, with 
$w\ne -1$ and in general varying with $z$, have been proposed 
(e.g., quintessence, k--essence, Chaplygin gas, Braneworld models, etc.);
even models that allow $w<-1$ (e.g., Phantom energy) have been considered
\citep[see for recent reviews][]{Sahni04,Padma06}.

We need further observations to unveil the true nature of what we call DE.  
A strong effort is being done now for developing the
next--generation SNIa experiments \citep[see e.g.][]{LH05} aimed
mainly to reduce random uncertainties. However, it is also crucial to
reduce systematic uncertainties as well as to break model
degeneracies \citep[e.g.,][]{weller02, LH03,
nesseris04,Ghisellini05,paper2}. 
The two latter papers illustrate how some degeneracies in the
cosmological parameter space can be reduced if the standard candles
used to construct the HD span a wide $z$ range. In this sense, long GRBs 
have been proposed as a class of objects able to extend the
HD up to very high redshifts in a natural manner \citep[][]{schaefer,gglf04}.

Despite the large dispersion of the long GRB energetics
\citep{frail01,bloom03}, the discovery that their energetics correlate
with observable quantities like the peak energy \Ep\ (in $\nu L_\nu$) 
of the time--integrated prompt emission spectrum and the achromatic
``break--time'' in the afterglow ligth curve \citep{ggl04,LZ05,Nava05}, 
has been used to ``standardize'' it. This allowed to employ GRBs as truly 
cosmological tools \citep[Ghirlanda et al. 2004b,2006,][]{dai04, fgga05,XDL05}.  
\citet{paper1} have found a new GRB
correlation whose tightness, in the framework of the standard fireball
scenario, is explained by its scalar nature.
Since such new correlation has the same shape in the observer and in 
the comoving frame, the influence of the $\Gamma$ relativistic
factor on the observed scatter becomes negligible.
The correlation is based on prompt $\gamma-$ray observables only 
(besides the redshift), 
by--passing the need of measuring afterglow quantities as is the
case of the ``Ghirlanda" relation \citep{ggl04}. The quantities
involved in the new correlation are the isotropic peak luminosity
\Liso, \Ep, and the ``high signal'' timescale \tdur,
previously used to characterize the variability behavior of bursts. In
\citet{paper2} we have found that by varing the cosmology, the data 
points present a minimum scatter around their best fit line in 
correspondence of the so--called $\Lambda$ cold dark matter ($\Lambda$CDM)
concordance cosmology: a flat geometry Friedmann-Robertson-Walker-Lema\^itre 
model with the cosmological constant dominating today. 
This result shows that, indeed, the \relL\ relation can be used to 
derive cosmological constraints.

Due to the lack of low$-z$ GRBs for calibrating a given correlation
independently from cosmology, the use of a 
statistical approach to {\it jointly} calibrate the correlation and 
constrain the cosmological parameters is required. 
\citet[][2006b]{fgga05} have presented an 
iterative Bayesian method to deal with this, so--called, 'circularity problem'.  
The same method can be used for SNIa.  Note that SNIa are not {\it
perfect} standard candles: their luminosities vary with the shape of
the light--curve (the brighter--slower relation) and with color (the
brighter--bluer relation) \citep[e.g.,][and the references
therein]{Guy05}. Due to several high$-z$ systematical effects, a
better calibration of these relations is obtained if higher$-z$ SNe
are included. The latter makes these relations cosmology
dependent. Therefore, the best fit to these relations has to be
carried out {\it jointly} within the same cosmological fit
\citep[][hereafter A05]{Astier05}. This approach has been applied to the 'Supernova
Legacy Survey' (SNLS) of 115 SNIa (A05). We apply here our
Bayesian method to this sample
to improve the constraints given by A05, who used a simple
multi--parametric \chs\ minimization method.

Observations of SNIa are accurate and the current samples 
comprise more than one hundred objects, but they are detected only at 
relatively low $z'$s, which introduces the degeneracy problem 
mentioned above. The GRBs useful as distance indicators span a 
large redshift range (up to $z=4.5$) but they are still scarce (19 bursts for the 
\relL\ relation).  Thus, a promising strategy to partially overcome the 
problems that each family of objects individually suffer of, 
is to combine both in the same Hubble 
diagram and use them jointly for constraining the cosmological parameters.
This is the goal of this Letter. Our main result is that the concordance
$\Lambda$ cosmology (minimal DE model) is fully consistent with the joint
GRB and SNLS SNIa data spanning the redshift range from $z=0.015$ to 4.5. 
Previous results with the so--called SNIa 'gold-set' ($z<1.7$) showed a
marginal inconsistency with the concordance model 
\citep[][]{Riess04,alam04,CP05,jassal05,NP05}, suggesting the possibility of
alternative DE models.  
 
In \S 2 we describe the SNIa and GRB samples used here. Section 3 deals
with the method while the results are presented in \S 4. The conclusions
of this work are given in \S 5.  

%========================================================
\section{The samples}
%========================================================

\subsection{Type Ia supernovae}

We analyze here the SNIa sample presented in A05. 
This sample consists of 44 nearby ($0.015<z<0.125$) SNIa assembled
from the literature, and of 73 distant SNIa {\bf ($0.249<z<1.010$)} 
discovered and followed during the first year of SNLS\footnote
{see \textit{cfht.hawaii.edu/SNLS/}}. 
The same light--curve fit method \citep{Guy05} was applied to 
all SNIa in the sample. 
For each SN, the reported quantities along with their errors are the 
fitted rest--frame $B-$band magnitude $m^*_B$ and the values of the 
parameters $s$ (light--curve stretch) and $c$ (normalized $B-V$ 
color at the maximum of the light--curve). 
The magnitude $m^*_B$ refers to observed brightness, and therefore 
does not account for the brighter--slower and the brighter-bluer 
correlations. As mentioned in the Introduction, A05 suggest to 
empirically calibrate these correlations using all objects (either at 
low or high$-z$). For the cosmological fits, two SNIa out of the 117 
SNLS sample objects were excluded because they are outliers in the HD. 

We follow A05 and include in the SNIa dispersions the contribution of 
a peculiar velocity of 300 km/s (this dispersion depends on $z$ and
on cosmology) and an intrinsic dispersion of SN absolute magnitudes of 0.132.
The latter is adjusted to obtain a SN reduced \chsr=1 and coincides with
the value of $0.13\pm 0.02$ given in A05 for their concordance cosmology.

\subsection{GRBs and the \relL\ relation}

The sample of 19 GRBs with known redshifts and with the observational 
information required to establish the $\Liso=CE_{\rm pk}^mT_{0.45}^{-n}$ 
correlation was presented in \citet{paper1}. 
The {\it rest frame} quantities entering 
in this correlation are the bolometric corrected (in the range of $1-10^4$ 
keV in the rest frame) isotropic--equivalent luminosity, \Liso, the spectrum 
peak energy, \Ep, and the time spanned by the brightest 45$\%$ of the total 
lightcurve counts above the background, \tdur. 
The assumed energy range for calculating \tdur\ is $50-300$ keV in 
the rest frame. 
We use the recipe proposed by \citet{Reichart01} to pass from the observed 
energy range to the assumed rest one, and the ligth-curve time binning of 
{\it HETE-II}, 164-ms \citep[see][]{paper1}. 

Besides the redshift, the observables required to estimate \Liso, \Ep\
and \tdur\ are: the peak flux $P$, the fluence $F$, the spectrum peak
energy \Ep, and its overall shape \citep[as described by the][model for most cases]
{Band93}, and the time scale of the brightest $45\%$ of the total light--curve 
counts above the background, \tdurob.  Errors on the observable parameters are
appropriately propagated to the composite quantities under the
assumption of no correlation among the measured errors.
For the concordance cosmology the correlation gives a reduced $\chsr=0.7$
\citep{paper1}, which could be signaling to an overestimate in the
data uncertainties. Whilst the observable uncertainties do not improve,
we can not attempt to estimate any intrinsic dispersion for our GRB sample.

%========================================================
\section{The method}
%========================================================

The  'circularity problem' mentioned in the Introduction in principle 
should not be a problem for SNIa because the brighter--slower and brighter--bluer 
correlations could be calibrated with a low$-z$ sample. However, the
distance estimates to high--$z$ SNIa improve if the parameters 
of these correlations are empirically determined along with the \chs\ 
minimization, from which the cosmological parameters are also constrained
(A05; note that multi--band photometric data are necessary 
to apply this technique). The situation is therefore, at least mathematically, 
similar to the circularity problem of GRBs. This suggests us to use our 
Bayesian method for improving the SNLS cosmological constraints estimated 
by A05. 

The basic idea of our approach is to find the best--fitted correlation on 
each point $\bar\Omega$ of the explored cosmological parameter space [for instance 
$\bar{\Omega} = (\bar{\Omega}_{\rm m},\bar{\Omega}_{\Lambda})$] and estimate 
with such correlation the scatter 
$\chi^2(\Omega,\bar{\Omega})$ on the HD for any given cosmology $\Omega$.
The conditional probability $P(\Omega|\bar{\Omega})$ inferred from the
$\chi^2(\Omega,\bar{\Omega})$ statistics provides the probability for each 
$\Omega$ given a possible $\bar{\Omega}$--defined correlation.
By defining with $P^\prime(\bar\Omega)$ an arbitrary probability for each
 $\bar\Omega$--defined correlation, the total probability of each $\Omega$,
using the Bayes formalism, is given by
\begin{equation}
P(\Omega) = \int P(\Omega|\bar{\Omega})P^\prime(\bar{\Omega})d\bar{\Omega},
\label{bayes}
\end{equation}
where the integral is extended on the available $\bar{\Omega}$ space.
Because the observations give a correlation for each cosmology, 
$P^\prime(\bar{\Omega})$ is actually the probability of the cosmology. 
Consequently such probability is obtained putting 
$P^\prime(\bar{\Omega}) = P(\Omega)$ and solving the integral Eq.(\ref{bayes}). 
This method based on the use of an iterative Monte Carlo technique has been
 presented in Firmani et al. (2005,2006b), and we refer the reader to such 
papers for more details. 

In what follows, we apply the Bayesian method to (i) the SNLS SNIa sample 
alone, and (ii) to the combination of both the SNIa and GRB samples.

%========================================================
\section{Results}
%========================================================

%--------------------------------------------------
\begin{figure}
\vskip -2.2 true cm
\hskip -1.7 true cm \psfig{file=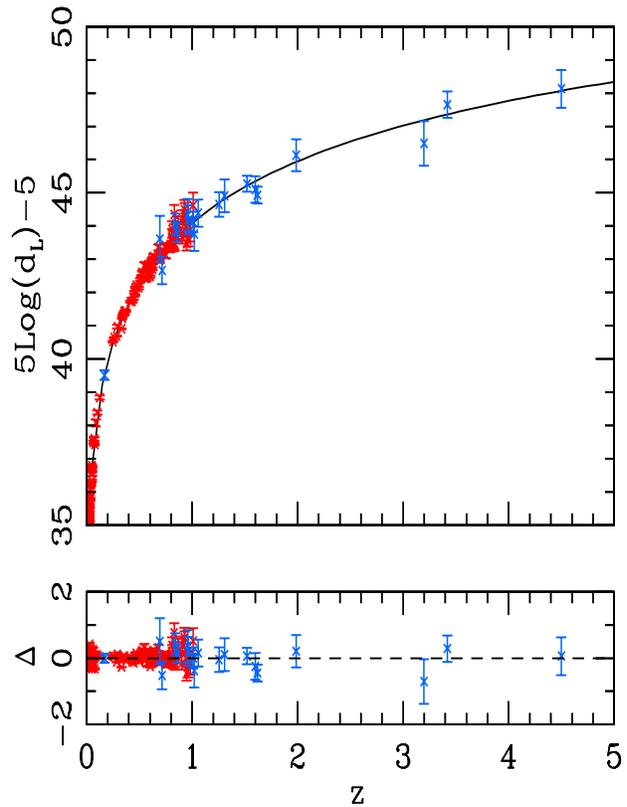,height=14.5cm,width=11.5cm}
\vskip -1.5 true cm
\caption{
Top panel: The SNIa (red symbols) and GRB (blue symbols)  
Hubble diagram (HD) for a concordance cosmology. 
In the bottom panel we show the residuals of the data points 
minus the concordance model.
}
\label{fig1}
\end{figure}

%--------------------------------------------------

Figure \ref{fig1} shows the HD assuming the concordance
cosmology (\Om=0.28, \OL=0.72 and h=0.71) for the 117 SNIa ($0.015<z<1.010$)
reported in A05 (red symbols), as well as for the 19 GRBs ($0.17<z<4.5$) from our
sample (blue symbols). From this plot one sees that GRBs are a
natural extension of SNIa to high redshifts.  The observational
uncertainties for GRBs are
still significantly larger than for SNIa.  The residuals of both
samples with respect to the assumed cosmology (solid curve) are shown
in the bottom panel of Fig. \ref{fig1}.  
The average of the absolute values of the residuals and its uncertainty
for the SNIa and GRB samples are $0.15\pm 0.01$ and $0.26\pm 0.05$, respectively.

% ----------------------------------------------

\begin{figure}
\vskip -2.5 true cm
\hskip -2 true cm \psfig{file=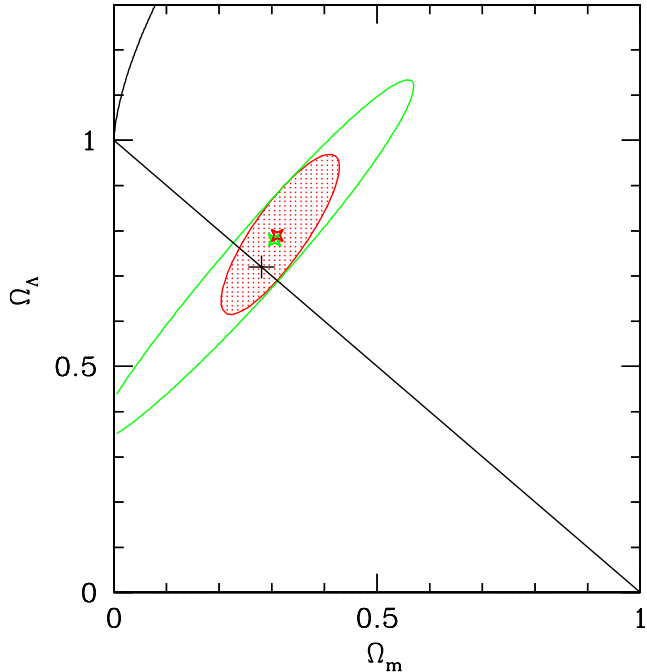,height=13.5cm,width=13.5cm}
\vskip -1.5 true cm
\caption{
Constraints on the (\Om,\OL) plane from the SNIa HD using our Bayesian 
approach to circumvent
the circularity problem (green line) and from the combined SNIa+GRB 
HD (red line and shaded region). Both lines are contours at 68.3\% CL's.
The star shows where \chsr\ reaches its minimum, while the cross indicates the
concordance cosmology.
The line corresponds to the flat geometry cosmology, the upper curve
is the loitering limit between Big Bang and no Bing Bang models 
}
\label{fig2}
\end{figure}

% ----------------------------------------------

In Fig. \ref{fig2} we show the 1$\sigma$ confidence levels (CL's) for only 
the SNIa sample (green line) and for the combined SNIa+GRB sample (red 
shaded region) in the \Om--\OL\ plane.  The use of the Bayesian method for
analyzing the first--year SNLS SNIa dataset improves somewhat the
cosmological constraints, especially for the large \OL\ part of the CL,
as can be appreciated by comparing Fig. \ref{fig2} with Fig. 5
in A05.  From Fig. \ref{fig2} we also see that the
inclusion of GRBs greatly improves the constraints given by SNIa
alone.  Because GRBs span a wide range of $z'$s, the degeneracy
between \OL\ and \Om\ is less severe for them than for the SNIa
\citep[see the discussion in][]{paper2}.  This achievement
is obtained despite the small number of GRBs and their relatively
large observational uncertainties.

Both the SNLS SNIa sample and the GRBs sample, show each one that 
the best--fit values of \Om\ and \OL\ are close to the flat--geometry case: 
the concordance model is actually well
inside the corresponding 1$\sigma$ CL constraints \citep[A05,][]{NP05,paper2}.
Now, the combined set makes the constrain even more restrictive.
If one forces \Ot = 1, our statistical analysis gives
$\Om = 0.273^{+0.027}_{-0.024}$.  
This range intersects the range of \Om\ values allowed by dynamical determinations
\citep[e.g.,][]{Hawkins03,schu03}.  
Thus the constraints to the $\Lambda$ cosmology parameters obtained
here are consistent with several other independent cosmological measurements.

\begin{figure}
\vskip -2.5 true cm
\hskip -2 true cm \psfig{file=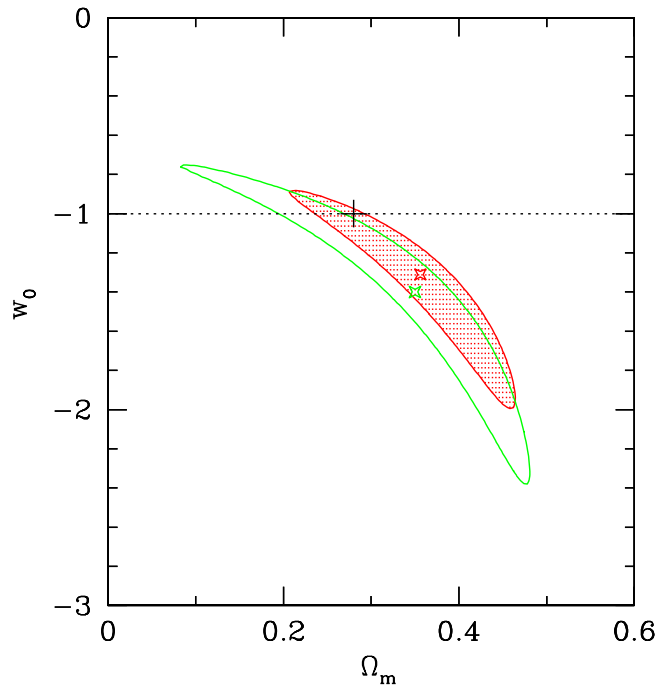,height=13.5cm,width=13.5cm}
\vskip -1.5 true cm
\caption{
Constraints on the (\Om, \wo) plane for a flat cosmology with static DE, 
using the same convention of Fig. \ref{fig2}. 
}
\label{fig3}
\end{figure}

\begin{figure}
\vskip -2.5 true cm
\hskip -2 true cm \psfig{file=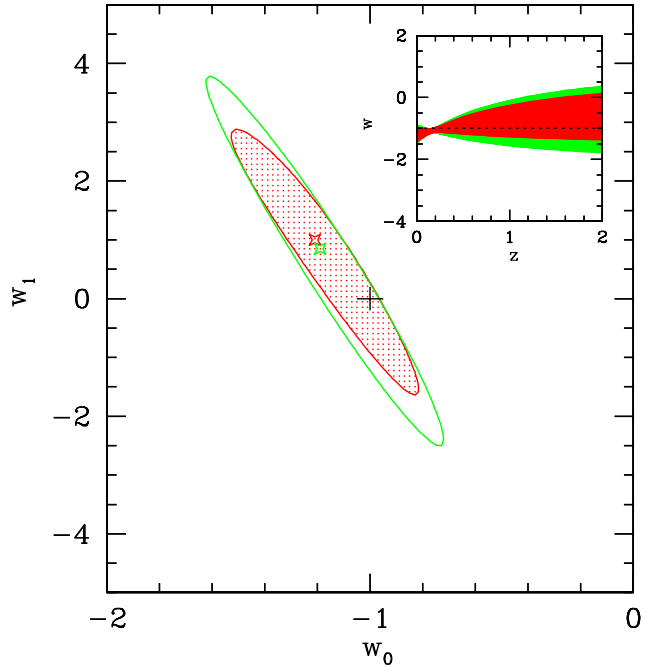,height=13.5cm,width=13.5cm}
\vskip -1.5 true cm
\caption{
Constraints on the (\wo,\wu) plane for a flat cosmology with dynamic DE,
\zt=1, and \Om =0.28, using the same convention of Fig. \ref{fig2}. 
The encapsulated panel shows the 1$\sigma$ CL for \wz corresponding to the 
SNIa dataset alone (green plus red region) and the SNIa+GRB (red region). 
} 
\label{fig4}
\end{figure}

\subsection{Flat cosmology with alternative DE}

We now relax the assumption $w=-1$, which was implicit up to now, and 
explore the possibility of $w=\wo$, where $\wo$ is a free parameter.
The limited number of objects in our two samples and the current
accuracies do not allow to have more than two free parameters. Therefore, we 
fix \Om+\OL=1, and find the CL contours in the $\wo-\Om$ plane.
Figure \ref{fig3} shows the 1$\sigma$ CL's in the $\wo-\Om$  plane
using only the SNLS SNIa sample (green line), and the combined 
SNIa+GRB sample (red line and shaded region).
Again, the SNIa constraints obtained with our iterative Bayesian method 
are tighter than the ones obtained in A05 (compare Fig. \ref{fig3} 
with their Fig. 5). 
When using the combined SNIa+GRB sample, we obtain a tight constraint
on \wo\ for reliable values of \Om.
For values of \Om\ in the range 0.236--0.286, the $\Lambda$ model ($w=-1$) 
is consistent at the  1$\sigma$. 
Assuming the prior \Om = 0.28, we obtain  $\wo=-1.055^{+0.073}_{-0.029}$,
while for \Om = 0.26, we obtain  $\wo=-1.000^{+0.055}_{-0.073}$.  
By combining independent cosmological probes that are sensitive to \Om~ 
\citep[e.g.,][]{Allen04, Eisen05} with our joint GRB+SNIa probe,
better constraints on \wo\ could be obtained. 

In order to explore the constraints on a possible evolution of $w$, based on 
the same arguments given in \citet{paper2}, we use the parametrization  
\citep{raw05} 
\begin{equation} 
\wz = \wo + \wu \frac{z} {\zt + z}.
\label{w} 
\end{equation}
Constraints on (\wo,\wu) for flat geometry {\it dynamical} DE models with the 
priors \Om=0.28 and \zt=1 are plotted in Fig. \ref{fig4}.  As discussed in 
\citet{paper2}, current
observational data do not allow yet to determine well \zt\  
\citep[$\zt=1$ corresponds to a popular parametrization introduced by][]{CP01}. 
The green line and the red shaded region represent the 1$\sigma$ CL's for the 
SNIa dataset alone and for the combined SNIa+GRB sample, respectively. 
The encapsulated panel shows the corresponding 1$\sigma$ CL for the 
evolution of \wz\ using the SNIa dataset alone (green plus red region) 
and the SNIa+GRB datasets (red shaded region).   
Again the $\Lambda$ case (\wo=$-1$ and \wu=0, which reduces to the concordance 
model because of the assumption that \Om = 0.28), is within the 1$\sigma$ CL's for
both SNIa and SNIa+GRB constraints. Our results allow at the 1$\sigma$ CL for 
models that avoid crossing the phantom dividing line ($w[z]=-1$). Notice that
although Eq. (\ref{w}) describes the evolution of $w$ up to any arbitrary 
large $z$ once its parameters were determined, the changes on $w$ with $z$ 
suggested by the observational constraints are formally valid only within 
the redshift range of the observational data, $z<4.5$ in our case.

\vspace{0.4cm}

%========================================================
\section{Conclusions}
%========================================================

We have combined a sample of potential standard candles that includes 
115 SNIa of the SNLS dataset (A05) and 19 long GRBs.  The latter were 
``standarized''  on the basis of a tight correlation among prompt 
$\gamma-$ray properties alone \citep{paper1}.  
Exploiting some similarities between the energetic callibrations of the SNLS 
SNe and the GRBs, we use for both populations of objects the same method to 
constrain cosmological parameters, namely a Bayesian approach described in 
Firmani et al. (2005,2006b). GRBs may be conceived as the natural extension 
of SNIa to large $z'$s in the HD. 
The advantage of this extension by using GRBs is that the heavy degeneracy 
in the space of the cosmological parameters due to a narrow and low $z$ range 
(as is the case of SNIa) is remarkably reduced. Our main results are as follow:

\begin{itemize}

\item
The cosmological constraints obtained with the Bayesian method for the SNLS 
sample alone are in agreement with those reported in A05. However, with the 
Bayesian method we obtain somewhat tighter constraints on \Om, \OL\ and \wo\
than with the \chs\ minimization method used in A05.
The values of \Om\ and \OL\ of the best fit (using SNIa only) are in good agreement 
with the flat--geometry case. Moreover, the best fitting values of \Om\ and \OL\ 
do not disagree with those obtained by using other cosmological probes 
\citep[e.g.,][]{spergel06}, re--affirming the reliability of the cosmological 
concordance model.

\item 
We have presented in a previous paper \citep{paper2} cosmological constraints 
derived by using GRBs only, whose energetics is standardized with the \relL\ 
relation.
We have shown here that the combined SNIa+GRB sample is able to largely reduce 
the allowed region of the parameter space with respect to the case where a
single population is used.
The resulting values of the best fits, once again, are in agreement with the 
concordance $\Lambda$CDM model. 

\item
As a consistency check, we have explored DE models with $w=\wo=$const (i.e. relaxing
the assumption of $w=-1$), but assuming flat geometry.
Furthermore, for completeness, we have allowed also $w$ to change with $z$ 
according to a parametric law and assuming \Om=0.28. In both cases we find 
that $w=-1=$const. is within the 1$\sigma$ CL from the best fits.

\end{itemize}

Finally, we emphasize the next two general conclusions: 

1) GRBs and SNIa as standard candles should not be considered as competing 
cosmological probes but as complementary methods.
Besides, both GRBs and high--$z$ SNIa suffer from the circularity 
problem concerning their calibration in such a way the same Bayesian method can be
applied for both samples.

2) According to our results there is no need for DE ``exotic" equations of state. 
The flat Friedmann-Robertson-Walker-Lema\^itre $\Lambda$ cosmology is fully
consistent with the HD constructed for the joint sample of SNIa and 
GRBs up to $z=4.5$. Similar conclusions were obtained recently on the basis of other
high--precision cosmological probes \citep{spergel06}.

\section*{Acknowledgments}

We acknowledge the anonymous referee for constructive comments
and Giuseppe Malaspina for technical support.
V.A-R. acknowledges the hospitality extended by INAF--OAB.
This work was supported by the italian INAF and MIUR 
(Cofin grant 2003020775\_002), and PAPIIT-UNAM grant IN107706-3.

\end{document}